\documentclass{birkart04}

 \theoremstyle{definition}

 \theoremstyle{remark}

 \numberwithin{equation}{section}

\begin{document}

%
%
%

%
\title[Electromigration of Islands with Crystal Anisotropy]
 {Islands in the Stream: \\ Electromigration-Driven Shape Evolution \\
with Crystal Anisotropy}
\author[Philipp Kuhn]{Philipp Kuhn}
\address{Fachbereich Physik,
Universit{\"a}t Duisburg-Essen\\
D-45117 Essen,
Germany}

\email{philipp@theo-phys.uni-essen.de}

\thanks{This work was supported by DFG within 
SFB 616 \textit{Energiedissipation an Oberfl\"achen.}}
\author{Joachim Krug}
\address{Institut f\"ur Theoretische Physik,
Universit{\"a}t zu K{\"o}ln\br
Z\"ulpicher Strasse 77,
50937 K{\"o}ln,
Germany}
\email{krug@thp.uni-koeln.de}


\date{March 25, 2004}

\begin{abstract}
We consider the shape evolution of two-dimensional islands on a crystal
surface in the regime where mass transport is exclusively along the 
island edge. A directed mass current due to surface electromigration causes
the island to migrate in the direction of the force. Stationary
shapes in the presence of an anisotropic edge mobility can be computed
analytically when the capillary effects of the line tension of the island edge
are neglected, and conditions for the existence of 
non-singular stationary shapes can be formulated. 
In particular, we analyse the dependence of the direction of island
migration on the relative orientation of the electric field to the
crystal anisotropy, and we show that no stationary shapes exist when the number
of symmetry axes is odd. 
The full problem including
line tension is solved by time-dependent numerical integration of the 
sharp-interface model. In addition to stationary shapes and shape instability
leading to island breakup, we also find a regime where the shape displays
periodic oscillations.

\end{abstract}

\maketitle

\section{Introduction}

The interest of researchers in the phenomenon of electromigration has in the past mainly been guided by the 
technological implications of this effect, namely the degeneration of integrated circuits 
\cite{Thompson93,Tu03,Suo03}. 
A large body of experimental work has been devoted to the study of electromigration-induced failure
of metallic conductor lines under accelerated testing conditions \cite{Tu03,Arzt94,Kraft95,Joo98}. At least for
simple geometries (e.g., a single void in a single crystal grain), such experiments have been successfully
modeled using a macroscopic continuum theory of shape evolution \cite{Kraft95,Gungor99}
(see Sect.\ref{Continuum}). On the other hand,
theorists have conducted a lively (and still ongoing) debate on the microscopic driving force of 
electromigration \cite{Verbruggen88,Sorbello98}. 

What appears to be largely missing in the field is experimental and theoretical work on the mesocopic
level, addressing the evolution of simple atomic-scale structures such as individual steps
and single-layer islands. This could help to bridge
the gap between the elementary moves of single atoms and the failure behavior of 
polycrystalline conductor lines with a complex internal geometry. In the present paper we address
the electromigration-induced shape evolution of single, atomic-height islands on a crystal
surface. We work in a continuum setting first introduced, in this context, by Pierre-Louis and 
Einstein \cite{PierreLouis00}. Among the different kinetic regimes considered in their work,
we focus here on the particularly simple case where mass transport is exclusively along
the island edge, so that the area of the island is strictly conserved and the evolution
law for the island boundary is local. Our new contribution consists in including crystal
anisotropy, which is clearly necessary to make contact with island electromigration 
on real crystal surfaces (such as silicon \cite{Metois99,Saul02}) and in lattice
Monte Carlo simulations \cite{PierreLouis00,Mehl00,Rous01}. 

The continuum model is introduced in the next section, and previous work on the problem is described.
In Section \ref{CapillarityFree} we show how stationary shapes can be explicitly computed when
the smoothening effects of the edge free energy is neglected, and numerical results for the full,
time-dependent problem are presented in Section \ref{Numerics}.   

\section{Continuum Model of Shape Evolution}
\label{Continuum}

The physical system consists of an island of monoatomic height on an otherwise flat crystal surface in 
interaction with the flow of the electrons. The island should be large enough to allow for a coarse grained 
description where the individual particles of the assembly are blurred into a continuous entity. To put this 
into a mathematical setting we represent the spatial conformation by a closed curve in the $xy$-plane 
parametrised by the arclength $s$. All geometrical and physical quantities can now be expressed as functions 
of $s$. We describe the local orientation of the island edge by the angle $\theta$ between the normal and 
the $y$-axis (counted positive in the clockwise direction). The derivative of $\theta$ with respect to $s$ 
is then the local curvature $\kappa$. The force $F$ acting on an atom at 
the island edge is composed of two contributions:
\begin{equation}
\label{Force}
F =  -\frac{\partial}{\partial s} (\tilde{\gamma} \kappa) + F_{el}.
\end{equation}
Here  $\tilde{\gamma} \kappa$ is the chemical potential of the island edge\footnote{Throughout we work
in units where the atomic area and the lattice spacing are set to unity.}, 
with $\tilde \gamma$ denoting
the edge stiffness, which is related to the step free energy $\gamma(\theta)$ by 
$\tilde \gamma = \gamma + d^2 \gamma/d \theta^2$ \cite{Krug04}. 
The gradient of the chemical potential on the right hand side of (\ref{Force}) 
accounts for the effect of capillarity which tends to minimize the free energy by driving material away from 
regions of high curvature. The second term is the force that actually causes electromigration.
It is conventionally written in the form $F_{el} = e Z^\ast E_t$, where $e$ is the elementary charge,
$E_t$ is the component of the local electric field which acts tangentially to the step edge,
and $Z^\ast$ is the effective valence of the step atom. On this level of modeling, all microscopic
aspects of electromigration are lumped into the quantity $Z^\ast$. 
To give an example, the effective valence for a copper atom moving along a close-packed step
on the Cu(100) surface has been calculated to be $Z^\ast \approx -31$ \cite{Rous01}. 
The form of the tangential electric
field to be used in this work will be specified later. 

The motion of the atoms along the edge can now be described by a mass current $j$ that is proportional to 
the acting force, with the factor of proportionality defining the orientation-dependent 
step edge mobility $\sigma (\theta)$,
\begin{equation}
j = \sigma(\theta) F.
\end{equation}
If we assume that atoms can neither detach from the island nor attach to it, the equation of shape evolution
 is simply the continuity equation that relates the divergence of the mass current to the normal velocity 
$v_n$ of the island edge, as
\begin{equation} \label{eqofmotion}
v_n = -\frac{\partial}{\partial s} j = \frac{\partial}{\partial s} \sigma (\theta) [ 
\frac{\partial}{\partial s} (\tilde{\gamma} \kappa) - e Z^\ast E_t].
\end{equation}
By dimensional analysis, 
the comparison of the two terms inside the square brackets on the right hand side of (\ref{eqofmotion})
defines the characteristic length scale \cite{PierreLouis00,schimschakkrug,suowangyang}
\begin{equation}
\label{lE}
l_E = \sqrt{\frac{\tilde \gamma}{\vert e Z^\ast E_0 \vert}},
\end{equation}
where $E_0$ is a typical value of the electric field strength.
For islands small compared to $l_E$ the capillary forces dominate the evolution, while for islands
large compared to $l_E$ the evolution is dominated by the electromigration force. 
Thus $l_E$ sets the length scale on which electromigration-induced shape instabilities can
be expected.

As formulated so far, the model is identical to the one that has been used extensively to model the evolution
of quasi-two-dimensional (cylindrical) voids in current-carrying metallic thin films
\cite{Kraft95,Gungor99,ho,wangsuohao,Mahadevan96,Xia97,Schimschak98,Mahadevan99,Li99,schimschakkrug,
BenAmar00,Bhate00,Cummings01,Kim03}. In that context it is essential to take into account the effect of 
\textit{current crowding}, which refers to the fact that current lines cannot pass through the
insulating void, and hence the electric field is enhanced in the remaining parts of the conductor.
As it is necessary to monitor the changes in the 
electric current configuration that occur in response to the shape changes of the void,
the dynamical evolution becomes manifestly nonlocal. In the physical situation of interest in the
present paper, this complication does not arise, since a single-layer island is not expected
to significantly modify the electric current configuration in the underlying 
substrate film.
We may therefore assume a constant electric field of strength $E_0$ directed at an angle 
$\phi$ with respect to the $x$-axis, and set $E_t = E_0 \cos(\theta + \phi)$ in 
(\ref{eqofmotion}).  

A model that interpolates between the case of a completely insulating
void and the case of a constant electric field considered here is obtained by assigning a
conductivity $\Sigma'$ to the void which differs from the conductivity $\Sigma$ of the bulk
material \cite{ho,wangsuohao}. The two limiting cases then correspond to 
$\Sigma'/\Sigma = 0$ and $\Sigma'/\Sigma = 1$.
The local model with $\Sigma'/\Sigma = 1$ has also been used to describe the electromigration-driven
evolution of dislocation loops \cite{Suo94,yangwangsuo}. 
 
Our primary aim in this paper 
will be to determine the stationary shapes of the drifting island. 
The condition for an island to move at constant velocity $V$ along the $x$-axis
without changing its shape is
\begin{equation} \label{statcond}
v_n = V \sin (\theta).
\end{equation}
It has been shown long ago that in the absence of anisotropy ($\tilde \gamma$ and $\sigma$
independent of $\theta$)  a circle is a stationary solution of (\ref{eqofmotion}), for general
conductivity ratio $\Sigma'/\Sigma$ \cite{ho}.  
In the following we try to find corresponding results for the anisotropic case.

\section{Analysis in the Absence of Capillarity} 
\label{CapillarityFree}

Combining the equation of motion (\ref{eqofmotion}) and the stationarity condition (\ref{statcond}) gives
\begin{equation} \label{stateq}
\frac{\partial}{\partial s} \sigma (\theta) [\frac{\partial}{\partial s} (\tilde{\gamma} \kappa) - 
F_0  \cos (\theta + \phi)] = V \sin (\theta)
\end{equation}
with $F_0 = e Z^\ast E_0$. 
This is a nonlinear ordinary differential equation of fourth order for the shape, or alternatively of 
second order, when viewed as an equation for the curvature as a function of $\theta$ 
(using $\frac{\partial}{\partial s} = \kappa \frac{\partial}{\partial \theta}$). 
In this section the capillarity term proportional to the stiffness $\tilde \gamma$ will be neglected.
While it is true that the effects of capillarity become small for large islands (island radius
$\gg l_E$), it can hardly be justified to drop this term altogether, because it contains the highest
order derivative in the equation. In fact islands that are large compared to $l_E$ break
apart, rather than reaching a steadily drifting shape \cite{PierreLouis00} (see Sect.\ref{Numerics}). 

Nevertheless the
analysis of the capillarity-free case $\tilde \gamma = 0$ is rather instructive.
It is accessible to an analytical treatment, because the unknown curvature enters (\ref{stateq}) only 
through the derivative $\frac{\partial}{\partial s}$ which can be eliminated by the following substitution: 
From the two geometrical relations
\begin{equation}
(ds)^2 = (dx)^2 + (dy)^2 \Leftrightarrow \left( \frac{\partial x}{\partial s} \right)^2 + \left( \frac{\partial y}{\partial s} \right)^2 = 1
\end{equation}
and
\begin{equation} \label{eq1}
\tan (\theta) = -\frac{dy}{dx} = -\frac{\;\frac{\partial y}{\partial s}\;}{\frac{\partial x}{\partial s}}
\end{equation}
we conclude that $\frac{\partial y}{\partial s} = - \sin (\theta)$. Replacing $\sin(\theta)$ in (\ref{stateq}) 
using this relation we find that the derivative of $V y-j$ is zero and therefore 
\cite{Suo94} $V y = j + const$. 
The constant of integration can be set to zero, since it generates only a vertical displacement. 
Hence the $y$-coordinate of the island shape as a function of $\theta$ is 
\begin{equation} \label{shapey}
y(\theta) = \frac{F_0}{V} \sigma(\theta) \cos (\theta + \phi)
\end{equation}
and using $\frac{dy}{dx} = - \tan(\theta)$ once more, we obtain
\begin{equation} \label{shapex}
x(\theta) = - \int_{0}^{\theta} \cot(\theta') \frac{\partial y}{\partial \theta'} \, d\theta'.
\end{equation}

In order to derive stationary island shapes from (\ref{shapey}) and (\ref{shapex}) 
one has to specify the mobility $\sigma (\theta)$. 
To mimic the anisotropy of a crystal surface with a certain discrete 
rotational symmetry, we choose \cite{schimschakkrug} 
\begin{equation}
\label{sigma}
\sigma (\theta) = \frac{\sigma_0[1+S \cos^2(D(\theta + \alpha))]}{1+S \cos^2(D \alpha)}
\end{equation}
where $\alpha$ is the angle between the symmetry axes and the coordinate 
system\footnote{The reason why we do not choose a symmetry axis to coincide with one of 
the coordinate axes is that the stationarity condition (\ref{statcond}) becomes particularly 
simple when the direction of motion coincides with the $x$-direction.},
$2D$ is the number of symmetry axes, and $S \geq 0$ is a measure for the strength
of the anisotropy; see Fig.\ref{AnisoFig} for an illustration of (\ref{sigma}). 
The denominator ensures that the mobility at $\theta = 0$ is $\sigma_0$. 

\subsection{Existence of stationary shapes}

When calculating the shape by means of (\ref{shapey}) and (\ref{shapex}) with a given choice 
of $D,S,\alpha$ and $\phi$, one encounters two types of irregularities: (i) The curve is generally 
not closed, and (ii) it can contain self-intersections. To get a closed shape one has to 
demand that $x$ and $y$ are $2 \pi$-periodic functions of $\theta$. 
Because $y(\theta)$ is always $2 \pi$-periodic it suffices to have
\begin{equation} \label{closenesscond}
\int_0^{2 \pi} dx(\theta) = \int_0^{2 \pi} d\theta \; \cot(\theta) [\frac{d \sigma}{d \theta} 
\cos(\theta + \phi) - \sigma \sin(\theta + \phi)] = 0.
\end{equation} 
For integer values of $D$ (even rotational symmetry) the mobility is $\pi$-periodic and hence the  
integrand in (\ref{closenesscond}) is odd under shifts $\theta \to \theta + \pi$.
As a consequence  
(\ref{closenesscond}) is always fulfilled. For half-integer $D$ (odd rotational symmetry) the 
evaluation of the 
integral in (\ref{closenesscond}) yields
$2D \cos (\phi) \cos(2D\alpha) - \sin (\phi) \sin(2D\alpha)$ 
times a nonzero factor, so that (\ref{closenesscond}) requires that 
\begin{equation} \label{closenesscond2}
\tan (2D\alpha) \tan (\phi) = 2D.
\end{equation}

\begin{figure}[htb]
\begin{center}
\includegraphics[width=0.25\textwidth,angle=-90,keepaspectratio]{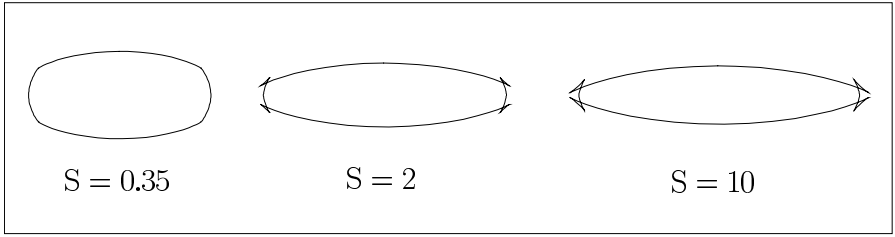}
\end{center}
\caption[]{Formation of self-intersections in the capillarity-free case. The figure shows
stationary shapes for $D=3$ and $\alpha - \phi = 0$, where the critical anisotropy strength is
$S_c \approx 0.35.$}
\label{Self}
\end{figure}

The generation of self-intersections is illustrated in Fig.\ref{Self}. 
Because the self-intersections are always accompanied by singularities (sign changes)
of the curvature, they can be avoided by requiring that
\begin{equation} \label{smoothnesscond}
\frac{1}{\kappa} =  -\frac{1}{\sin(\theta)} \frac{\partial y}{\partial \theta} > 0
\end{equation}
everywhere.
Thus the smoothness condition is related to the existence of extrema of $y(\theta)$. 
If $\frac{\partial y}{\partial \theta}$ is nonzero at $0$ or $\pi$, the integral in (\ref{shapex}) 
diverges because it contains the singularity of the cotangent function; the shape is then not bounded. 
If $\frac{\partial y}{\partial \theta}$ is zero somewhere else (\ref{smoothnesscond}) is violated. 
If $D$ is half-integer, the condition 
$\frac{\partial y}{\partial \theta}(0)=\frac{\partial y}{\partial \theta}(\pi)=0$ requires $\alpha = \phi = 0$, 
which violates (\ref{closenesscond2}). In other words: For half-integer values of $D$ the shape is either 
not closed or not bounded. 
While this result has been derived using the specific functional form (\ref{sigma}) for the
mobility, we expect it to be true in general that
\textit{in the capillarity-free case, no stationary shapes exist when the number
of symmetry axes is odd.}  

\begin{figure}[htb]
\begin{center}
\includegraphics[width=0.6\textwidth,angle=90,keepaspectratio]{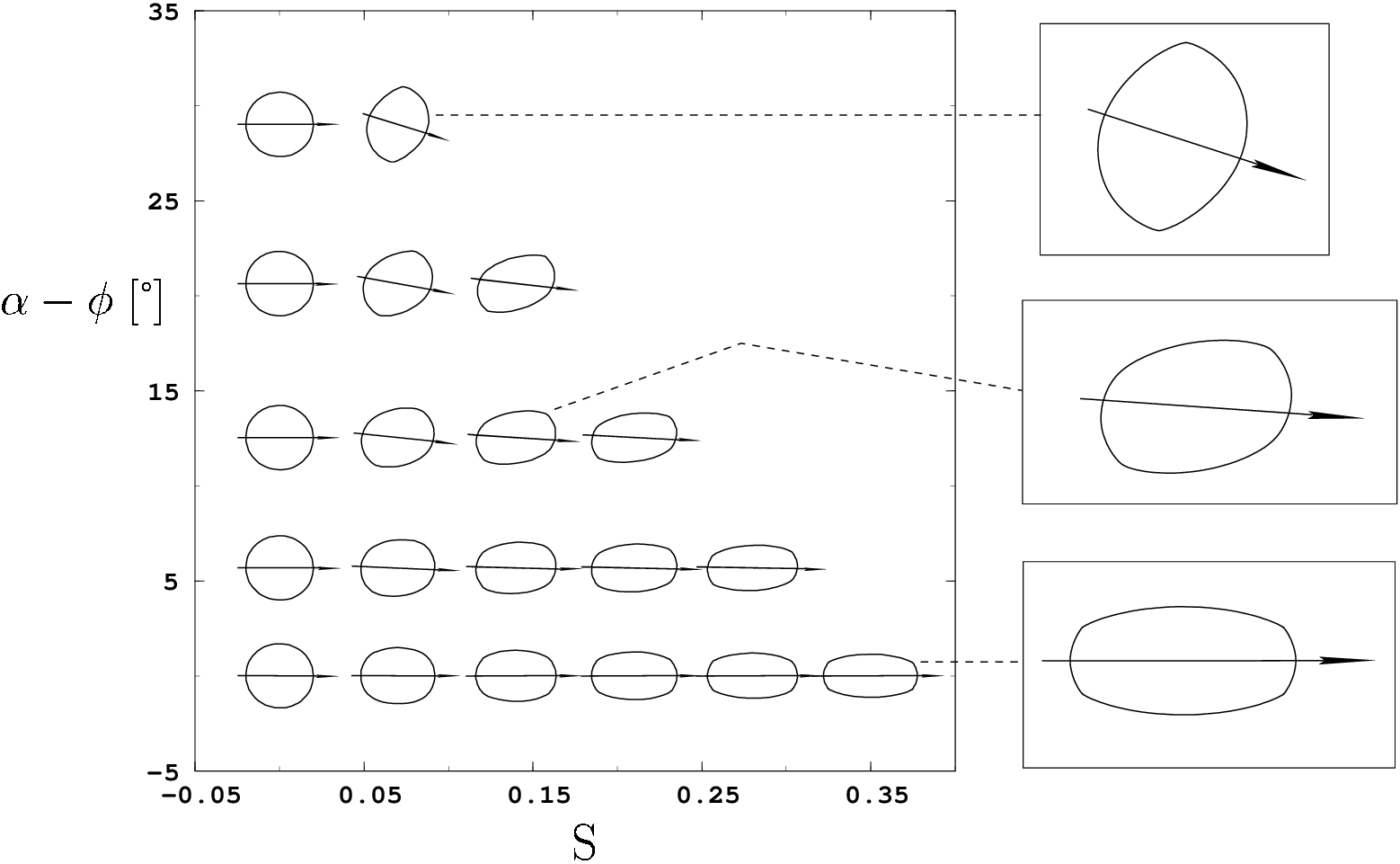}
\end{center}
\caption[]{Stationary island shapes without capillarity, for sixfold crystal anisotropy and
different values of the
angle $\alpha - \phi$ between the symmetry axis of the anisotropy and the field,
and the anisotropy strength $S$. In this figure the field direction is
horizontal, and the arrows indicate the direction of island migration. In the empty part
of the $(S,\alpha-\phi)$-plane no stationary shapes exist.}
\label{FreeFig}
\end{figure}

In the following
we may therefore restrict ourselves to anisotropies of even symmetry. 
In this case $\frac{\partial y}{\partial \theta} (0) =\frac{\partial y}{\partial \theta}(\pi)$
because of the $\pi$-periodicity of $y$ and we end up with two conditions, 
\begin{eqnarray}
\frac{\partial y}{\partial \theta} (0) = 0 &\Leftrightarrow& \frac{SD \sin(2D\alpha)}{1+S\cos^2(D\alpha)} = -\tan(\phi) \label{cond1}\\
\frac{\partial y}{\partial \theta} (\theta) \neq 0 &\Leftrightarrow& \frac{SD \sin(2D(\theta+\alpha))}{1+S\cos^2(D(\theta + \alpha))} \neq -\tan(\theta+\phi) \;\; \mbox{if} \;\;\theta \neq 0,\pi. 
\label{cond2}
\end{eqnarray}  
For a given value of the 
angle $\alpha - \phi$ between the symmetry axes and the field, (\ref{cond1}) determines the direction of motion 
of the island, and the inequality (\ref{cond2}) 
determines the values of $S$ for which a smooth shape exists. 
For $S=0$ there is always such a shape for every choice of $\alpha - \phi$, namely the circle 
which moves in the direction of the applied field. Increasing the anisotropy strength
$S$ the circle elongates along the field direction and finally at some critical value $S_c$ 
sharp corners develop as precursors of self-intersections (see Fig.\ref{Self}). 
For $S > S_c$ no stationary shapes exist.
Figure \ref{FreeFig} shows the array of possible shapes for the case of sixfold symmetry 
($D = 3$). Note that in the absence of capillarity the characteristic scale (\ref{lE}) disappears
from the problem, and hence the shapes are independent of island size (the island size determines
however the migration speed, see Sect. \ref{Speed}).

It is straightforward to derive the critical anisotropy strength $S_c$ for the special case
$\alpha = \pi/2D$, $\phi = 0$, where the direction of island motion and the field direction
coincide with the direction of \textit{minimal} mobility. The island then elongates perpendicular
to the field and the self-intersections are forced by symmetry to appear at $\theta = 0$ and 
$\pi$. The condition of vanishing curvature radius reads simply $d^2 y/d \theta^2 \vert_{\theta=0} = 0$, 
which implies
$[\sigma- d^2 \sigma/d\theta^2] \vert_{\theta=0} = 0$. 
For our choice (\ref{sigma}) of the mobility function
this yields 
\begin{equation}
\label{Sc0}
S_c = \frac{1}{2 D^2 - 1} \;\;\; \textrm{for} \;\;\; \alpha = \frac{\pi}{2D}, \; \phi = 0.
\end{equation}
This is the minimal value of $S_c$; the maximum range of smooth shapes appears at
$\alpha = \phi = 0$, where the island moves along the direction of maximal mobility (Fig.\ref{FreeFig}).

\subsection{Direction of island migration}
\label{Direction}

An important consequence of anisotropy, which remains true also when capillary forces
are turned on, is that the direction of island motion does not generally coincide with the
direction of the electromigration force when $\alpha - \phi \neq 0$. This effect was
previously observed for the nonlocal model \cite{schimschakkrug}. The relationship 
(\ref{cond1}) shows that $\alpha$ and $\phi$ have opposite signs, which implies that
the direction of island motion lies between the field direction and the symmetry axis.

\begin{figure}[htb]
\begin{center}
\includegraphics[width=0.8\textwidth,keepaspectratio]{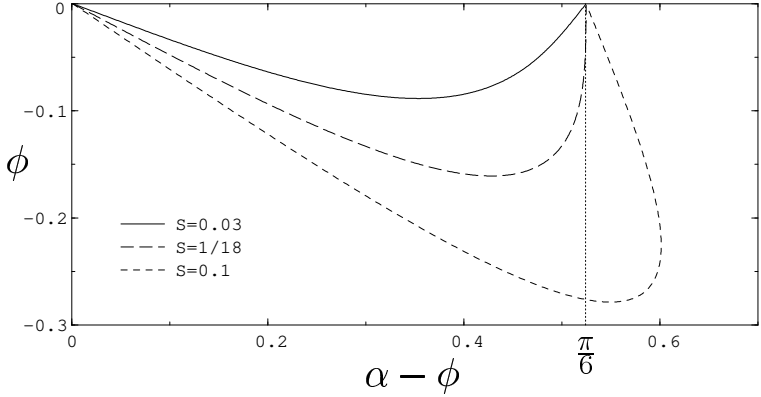}
\end{center}
\caption[]{Relative angle $\phi$ between the field direction and the direction of island motion
as a function of the angle $\alpha - \phi$ between the field and the axis of maximal
mobility, as computed from (\ref{cond1}) with $D=3$. The vertical line marks the minimal
mobility direction $\alpha = \pi/6$.}
\label{Curves}
\end{figure}

Figure \ref{Curves} shows the angle between the direction of
island motion and the field, $\phi$, as a function of $\alpha - \phi$, which is the physical control
parameter determined by the experimental setup. The island moves in the direction of the field,
$\phi = 0$, both when the field is along the maximal mobility direction ($\alpha - \phi = 0$)
and along the minimal mobility direction ($\alpha - \phi = \pi/2D$). With increasing
anisotropy strength the graph becomes strongly skewed towards the minimal mobility direction until,
at a second critical anisotropy strength given by $\tilde S_c = 1/2 D^2$, the function becomes 
multi-valued. Beyond this point $\phi$ approaches a nonzero limit as $\alpha - \phi \to 
\pi/2D$, despite the fact that $\phi = 0$ exactly at $\alpha - \phi = \pi/2D$. The physical
consequence is that the direction of island motion, as well as the island shape, change
\emph{discontinuously} as the field direction is moved across the direction of minimal mobility.
Since $\tilde S_c$ is (slightly) smaller than the minimal value of $S_c$ given by (\ref{Sc0}),
there is a small range of anisotropy strengths where smooth shapes exist for all
angles but the shape changes
discontinuously at $\alpha - \phi = \pi/2D$.          

\subsection{Migration speed}
\label{Speed}

Since both coordinates of the parametrization (\ref{shapey}) and (\ref{shapex}) 
are multiplied by $1/V$ it is clear that the velocity is inversely
proportional the extension of the island; for a circular island of
radius $R$ it is well known that $V \sim 1/R$ \cite{ho}. 
In other words, for every allowed choice of $\alpha - \phi$ and $S$,
(\ref{shapey})  and (\ref{shapex}) generate a whole family of similar shapes
moving with different velocities. To determine the dependence of $V$ on the
island size we calculate the area $A$ of the island as follows:
\begin{eqnarray*}
A = \int_{0}^{2 \pi} y(\theta) \, dx(\theta) = - \frac{1}{2} \int_0^{2 \pi} d\theta
\; \cot(\theta) \frac{d}{d\theta} y(\theta)^2 \\ = 
- \frac{1}{2} \left( \frac{F_0}{V} \right)^2 \int_0^{2 \pi} d\theta
\; \cot(\theta) \frac{d}{d\theta} \sigma(\theta)^2 \cos^2(\theta + \phi).
\end{eqnarray*}
Inserting the form (\ref{sigma}) for the mobility, it follows that the
migration speed is of the form
\begin{equation}
\label{speed}
V = \frac{F_0 \sigma_0}{\sqrt{A}} \; \frac{\sqrt{a S^2 + b S + \pi}}{1 + S \cos^2(D \alpha)},
\end{equation}
where the coefficients $a$ and $b$ are integrals of combinations of trigonometric
functions which depend on $\alpha$, $\phi$ and $D$. For the special case 
$\alpha = \phi = 0$ and $D=3$ illustrated in the bottom row of Fig.\ref{FreeFig},
we find $a = 39 \pi/8$ and $b= 7 \pi$, which implies that the speed increases 
by about a factor of 1.5 when the anisotropy is increased from $S=0$ to the critical
value $S_c \approx 0.35$. The speed increase is related to the elongation of the island
shape, which brings the orientation of the edge closer to the maximum mobility
direction $\theta = 0$. When the island moves along the direction of minimal mobility
($\alpha = \pi/2 D, \phi = 0$) the coefficients in (\ref{speed}) are negative,
$a = -9 \pi/8$ and $b = - 5 \pi$, which implies that the island slows down with
increasing anisotropy.

\begin{figure}[htb]
\centerline{
\includegraphics[width=0.32\textwidth,angle=-90,keepaspectratio]{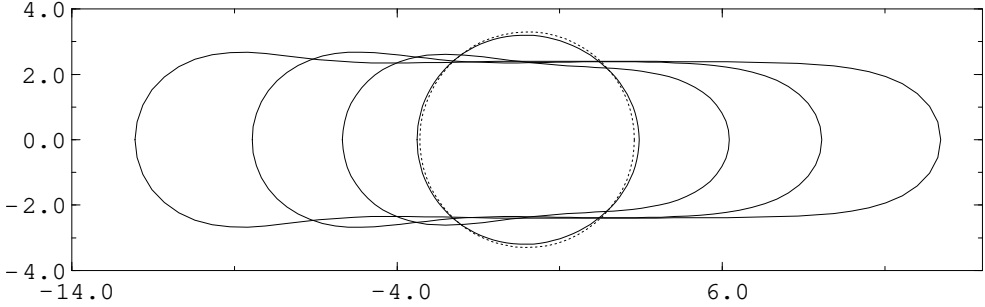}
}
\caption[]{Stationary shapes without anisotropy. The dimensionless 
initial radii of the different shapes are $\rho_0 = 3.3$, 4, 5 and 6, and the computation was done
with 100 discretization nodes. The dotted line shows the circular shape which constitutes the stable
stationary solution for $\rho_0 < \rho_c \approx 3.26$.}
\label{IsotropFig}
\end{figure}

\section{Numerical Integration of the Time-Dependent Problem}
\label{Numerics}

We now integrate the full dynamical problem (\ref{eqofmotion}) by means of the sharp interface method 
described in \cite{schimschakkrug,Schimschak99}. The evolution equations are discretized and integrated
using a variable-step variable-order predictor-corrector method \cite{Shampine75}. To avoid frequent
remeshing an additional tangential velocity was introduced which keeps the nodes equidistant and which
has to be computed implicitly.

Again we are mainly interested in the long time behavior of 
the system. As initial condition we take a circular island, usually imposed with a slight deformation to 
trigger a certain mode of motion\footnote{Unless stated otherwise, for the examples
described here the asymptotic mode of evolution is independent of the
precise initial condition.}. We restrict ourselves to the case of a mobility with six-fold symmetry ($D=3$)
with the field aligned parallel to a symmetry axis. The stiffness $\tilde \gamma$ is assumed
to be isotropic\footnote{This is physically motivated by the fact that the anisotropy of the mobility
$\sigma$, which is a thermally activated quantity, is expected to be much more pronounced than that of 
$\tilde \gamma$.}. 
The tunable parameters are then the initial radius $R_0$ and the strength $S$ of the anisotropy.
All lengths are expressed in units of the characteristic scale $l_E$, and the  
dimensionless initial radius is $\rho_0 = R_0/l_E$.  

\begin{figure}[htb]
\centerline{
\includegraphics[width=0.4\textwidth,angle=-90,keepaspectratio]{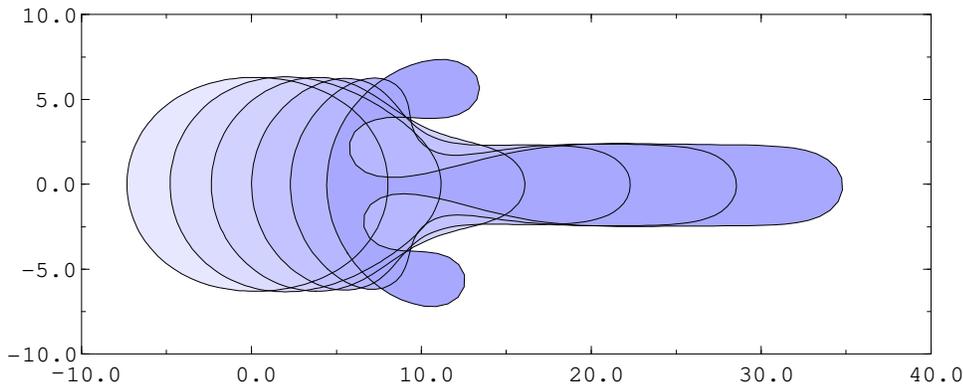}
}
\caption[]{Breakup of an island in the isotropic case. The figure shows a time sequence
of shapes, with a shading that darkens with increasing time. 
The initial shape is an ellipse elongated in the
direction of island motion, with an area corresponding to 
a circle of dimensionless radius $\rho_0 = 7$. 
The computation used 150 discretization nodes.
The breaking of the up-down symmetry is due to numerical noise.}
\label{IsotropBreak}
\end{figure}

To link our investigation to previous work we start with the isotropic case $S=0$. Here we find circular 
stationary solutions for small radii. At the critical dimensionless 
radius $\rho_c = 3.26$ the circular solution loses its
stability \cite{wangsuohao} and a bifurcation to non-circular stationary shapes occurs.
In \cite{yangwangsuo} this bifurcation was investigated by numerically solving the stationarity condition 
(\ref{stateq}), and two branches of post-bifurcation shapes were found. 
In our time-dependent integration we only observe one branch, referred to as `mode I' in \cite{yangwangsuo}.
To linear order in the distance from the bifurcation,  
the second `mode II' branch is the symmetrical counterpart with respect to the circle, in the sense that 
$\kappa_I(\theta) + \kappa_{II}(\theta) = const.$; mode I is elongated in the direction of the field, 
mode II perpendicular to it. The mode II shapes which move at a smaller velocity seem not to be selected by 
the dynamics.

Some noncircular stationary shapes for $S=0$ are shown in Fig.\ref{IsotropFig}. 
We see that for $\rho_0 \geq 3.5$ the shapes become non-convex. These shapes were not found in 
\cite{yangwangsuo}, because the numerical calculation was stopped at the point where $1/\kappa$ 
approaches zero. Increasing the dimensionless island size further above $\rho_0 \simeq 6.2$ 
causes the island to break. It first evolves a long finger, which corresponds to the `slit'-solution 
described in \cite{suowangyang}, and which subsequently detaches to form a new island
(Fig. \ref{IsotropBreak}). 
Similar breakups occur in the presence of anisotropy, so that the range of $\rho_0$ is generally 
restricted towards large values.

\begin{figure}[htb]
\centerline{
\includegraphics[width=0.5\textwidth,angle=-90,keepaspectratio]{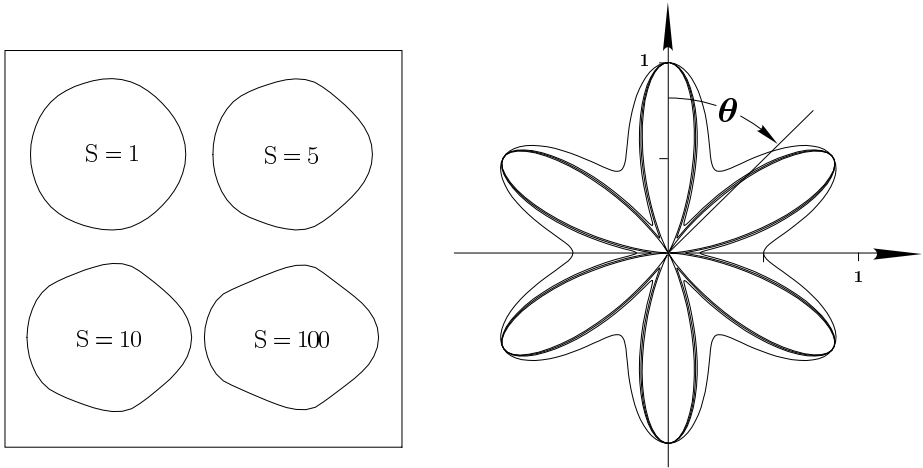}
}
\caption[]{Left panel: Stationary shapes with anisotropic mobility of sixfold symmetry ($D = 3$). 
The initial shape was a circle of dimensionless radius $\rho_0 = 2.5$, and the anisotropy
strength is $S=1$, 5, 10 and 100. The computation used 50 discretization nodes. 
Right panel: Polar plot of the mobility corresponding to the shapes in the left panel.}
\label{AnisoFig}
\end{figure}

For the anisotropic case there are no circular stationary shapes anymore. The deviation from the circle 
increases with increasing size $\rho_0$ and strength of anisotropy $S$. For strong anisotropy the islands 
evolve facet-like features, i.e. regions of the edge where the radius of curvature is very large. Figure 
\ref{AnisoFig} shows some examples. As was already discussed in Sect. \ref{Direction}, 
the direction of motion generally differs from the field direction, when the latter is not 
aligned with the symmetry axes of the anisotropy. Under certain conditions  
the symmetry of the direction of motion with respect to the field and anisotropy
direction appears to be \emph{spontaneously broken}, i.e. we have observed
stationary island migration off the field
direction even when the latter coincides with a 
symmetry axis of the anisotropy \cite{Kuhn04}. This behavior may be related to the
discontinuity of the direction of motion when the field direction crosses the direction
of minimal mobility, which was described in Sect.\ref{Direction} for the capillarity-free case.

\begin{figure}[htb]
\begin{center}
\includegraphics[width=0.27\textwidth,angle=-90,keepaspectratio]{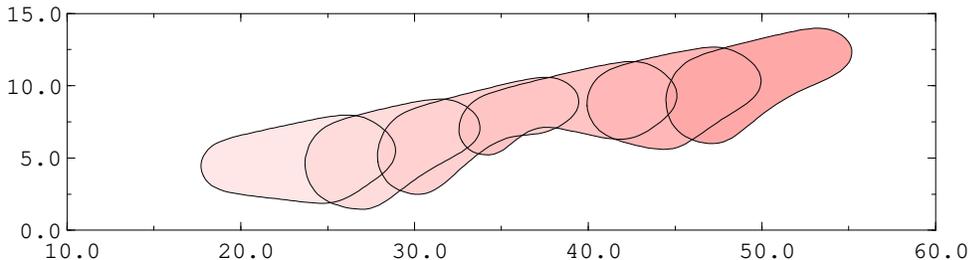}
\end{center}
\caption[]{Oscillatory island motion. The dimensionless initial radius was $\rho_0 = 4$ and
the anisotropy strength $S=1$. Note that the upper boundary is stable, while the lower
boundary supports a traveling wave. The computation used 100 discretization nodes.}
\label{OscFig}
\end{figure}

With increasing $S$ the range of $\rho_0$ for which stationary solutions exist shrinks toward 
smaller values and is replaced by a domain where a qualitatively new behavior appears. Here the time 
evolution neither runs into a steady state nor into a breakup event but instead becomes oscillatory. 
A typical sequence is shown in Fig.\ref{OscFig}: First the island elongates along the field direction, 
then one side of it bends so that the conformation becomes nonconvex. This leads to an acceleration
of the  mass transport which then shrinks the shape back to its original state. 
The oscillatory shape evolution in Fig.\ref{OscFig} is moreover seen to break the symmetry
with respect to the field and anisotropy direction (recall that the field is directed along the
$x$-axis which corresponds to a direction of maximal mobility, see the right panel in 
Fig.\ref{AnisoFig}). The solution shown in Fig.\ref{OscFig} coexists with one obtained by
reflection at the $x$-axis, in which the island travels to the right and down. Which of the two
solutions is realized depends on the initial conditions. For other parameter values we have
also observed solutions that switch between periods of upward and downward motion \cite{Kuhn04}.  

The conditions for the appearance of stationary and oscillatory shapes largely remain to be clarified.
On a qualitative level, we have observed that the  
``facets'' that dominate the stationary shapes 
for large $S$ tend to be close to orientations of maximal stability in the sense of linear stability
analysis \cite{Krug94}. In contrast, the oscillatory shapes are typically bounded by one 
stable and one unstable orientation, and the shape oscillation can be viewed as the propagation
of a wave along the unstable side of the island (see Fig.\ref{OscFig}).  

For the reasons described in Sect.\ref{CapillarityFree}, 
a quantitative comparison between the shapes computed numerically for the full problem and 
the analytically derived capillarity-free shapes is not meaningful. On a qualitative level,  
the most apparent difference between the shapes in Fig.\ref{FreeFig} and 
Figs.\ref{IsotropFig},\ref{AnisoFig} is that in the presence of capillarity the direction of motion can
clearly be discerned from the shape, while the capillarity-free shapes in Fig.\ref{FreeFig}
are front/back symmetric.
This can be seen directly from (\ref{stateq}). If $\tilde{\gamma} = 0$ the equation reduces to 
(again using $\frac{\partial}{\partial s} = \kappa \frac{\partial}{\partial \theta}$)
\begin{equation}
F \kappa \frac{d}{d \theta} \sigma (\theta)  \cos (\theta + \phi) = V \sin (\theta)
\end{equation}
which possesses two symmetries: First, reversing the direction of the field and the velocity simultaneously 
leaves the equation invariant. The second symmetry is the $\pi$-periodicity of the equation\footnote{The 
mobility is $\pi$-periodic only for integer $D$, which is also the reason why we always get closed forms in 
this case (see Sect.\ref{CapillarityFree}).} 
which results in periodicity of the solution. Both symmetries are destroyed for nonzero stiffness.

\section{Conclusions}

It is interesting to compare the local ($\Sigma'/\Sigma = 1$) and nonlocal ($\Sigma'/\Sigma < 1$)
boundary evolution problems in the light of our work\footnote{For a similar comparison
in the substrate geometry see \cite{Schimschak97}.}. For the isotropic case, it is known that
the circular solution becomes linearly unstable for sufficiently 
large radius when $\Sigma'/\Sigma > 0$ \cite{wangsuohao},
while it retains its stability (but becomes \textit{nonlinearly} unstable under a sufficiently
large perturbation \cite{Schimschak98,Cummings01}) when $\Sigma'/\Sigma = 0$ \cite{wangsuohao,Mahadevan96}.
Furthermore it has been shown by complex variable techniques that noncircular stationary shapes
cannot exist for $\Sigma'/\Sigma = 0$ \cite{Cummings01}, hence a void is forced to break up
when it becomes unstable \cite{Schimschak98}. Here we have shown that breakup occurs for 
sufficiently large islands also in the local case, but in addition there is an intermediate
regime, between the linear instability of the circular solution and breakup, where 
noncircular stationary solutions are stable\footnote{Note that in \cite{yangwangsuo} the stability
and dynamical relevance of the noncircular stationary solutions was not addressed.}.

Perhaps
our most remarkable finding is that crystal anisotropy may induce complex, oscillatory evolution
modes in the local problem. An example of oscillatory void evolution was previously reported by
Gungor and Maroudas in the nonlocal case \cite{Gungor00}. Future work should address the robustness
and generality of the phenomenon. For this purpose it should be useful to employ numerical methods
that are complementary to the sharp interface algorithm adopted here, such as 
finite element \cite{Xia97}, phase-field 
\cite{Mahadevan99,Bhate00,Kim03}, level set \cite{Li99} and
kinetic Monte Carlo \cite{PierreLouis00,Mehl00,Rous01}
approaches, and to extend the study of the additional kinetic regimes (other than the case of edge
diffusion considered here) in \cite{PierreLouis00} to include crystal anisotropy.



\subsection*{Acknowledgment}
We would like to thank Martin Schimschak for helpful correspondence.

\end{document}